\documentclass[sigconf, nonacm]{acmart}

\newif\ifanon
\makeatletter
\if@ACM@anonymous
\anontrue
\fi
\makeatother

\usepackage[shortlabels,inline]{enumitem}
\setlist{nolistsep} 
\setlist[itemize]{leftmargin=*} 

\usepackage{mathtools}

\graphicspath{{figures/}}

\def \questionname {Q}
\newcommand{\qref}[1]{\questionname\ref{#1}}
\newcounter{qcounter}
\newenvironment{question}[1]{\refstepcounter{qcounter}\textbf{(\questionname\theqcounter)} #1}{}
\usepackage{subfigure}

\newcommand{\eg}{e.g., }

\newcommand{\ie}{i.e., }

\newcommand{\figref}[1]{Fig.~\ref{#1}}    

\newcommand{\secref}[1]{\S\ref{#1}}

\copyrightyear{2020}
\acmYear{2020}
\setcopyright{acmcopyright}\acmConference[e-Energy'20]{The Eleventh ACM
International Conference on Future Energy Systems}{June 22--26, 2020}{Virtual Event,
Australia}
\acmBooktitle{The Eleventh ACM International Conference on Future Energy Systems
(e-Energy'20), June 22--26, 2020, Virtual Event, Australia}
\acmPrice{15.00}
\acmDOI{10.1145/3396851.3403509}
\acmISBN{978-1-4503-8009-6/20/06}

\newcommand{\tdiff}{\Delta t }
\newcommand{\ti}{t_{i} }
\newcommand{\til}{t_{i-1} }

\begin{document}

\title[Poster: Defining a synthetic data generator for realistic EV sessions]{Poster: Defining a synthetic data generator \\for realistic electric vehicle charging sessions}

\author[M. Lahariya]{Manu Lahariya}
\affiliation{%
  \institution{IDLab, Ghent University -- imec}}
\email{manu.lahariya@ugent.be}

\author[D. Benoit]{Dries Benoit}
\affiliation{%
  \institution{Center for Statistics, Ghent University}}
\email{dries.benoit@ugent.be}

\author[C. Develder]{Chris Develder}
\affiliation{%
  \institution{IDLab, Ghent University -- imec}}
\email{chris.develder@ugent.be}


\begin{abstract}
Electric vehicle (EV) charging stations have become prominent in electricity grids in the past years. Analysis of EV charging sessions is useful for flexibility analysis, load balancing, offering incentives to customers, etc. 
Yet, limited availability of such EV sessions' data hinders further development in these fields. 
Addressing this need for publicly available and realistic data, we develop a synthetic data generator (SDG) for EV charging sessions. Our SDG assumes the EV inter-arrival time to follow an exponential distribution. Departure times are modeled by defining a conditional probability density function (pdf) for connection times. This pdf for connection time and required energy is fitted by Gaussian mixture models. Since we train our SDG using a large real-world dataset, its output is realistic.

\end{abstract}



\maketitle

\section{Introduction}
\label{sec:Intro}

    With the increased use of electric vehicles (EVs) over the past decade, a large network of EV charging stations has been installed on the electric grid. 
    Data collected from city-wide deployment of EV charging stations can be used for both academic and industrial purposes, \eg through statistical analysis \cite{nasrinEV2016} and modeling. Studies require reliable sessions data for understanding behaviors and exploring flexibility.  The scarcity of reliable data has been discussed previously \cite{EVDSreview}, and its necessity has been pointed out for further research purposes. Even when data exists, it may be protected under confidentiality for private collectors, and not freely available for academic or public use. Lack of wide-spread availability and accessibility of realistic EV charging sessions data pose a significant hurdle, impeding further research in the field.
    
    We propose a Synthetic Data Generator (SDG), that can be used to generate a sample of EV charging sessions. This implies temporal statistical modeling of arrivals and modeling of departures and the associated electrical load (\ie required energy). We define a parametric model that can be trained from a real-world dataset. This trained model can be subsequently used to generate realistic data samples of EV sessions. 
    In this paper, we contribute with:
        \begin{itemize}[topsep=0pt]
            \setlength\itemsep{0em}
            \item An approach to model sessions data for EVs over a group of charging stations defined as the SDG (\secref{sec:SDG});
            \item An outline of the model parameters, and discussion over benefits and drawbacks of different models;
            \item Answers to our main research questions, being:
        	\begin{question}\label{q:SDGdefination}~Which parametric models can be used to describe sessions of EVs. What are the input parameters and latent variables for these models? \end{question}
        	\begin{question}\label{q:generate}~How can we generate synthetic samples of EV sessions data from these parametric models? \end{question}
        \end{itemize}

\section{Synthetic Data Generator}
\label{sec:SDG}

Each EV session can be described using three parameters:
\begin{enumerate*}[(i)]
        \item \textit{Arrival time},
        \item \textit{Departure time}, and 
        \item \textit{Energy charged} (in kWh).
\end{enumerate*}
We can define three models for each of these as:
\begin{itemize}[topsep=0pt]
    \item \textit{Arrivals} = $AM$(Horizon)
    \item \textit{Connected times} = $MM_{c}$(Arrivals) 
    \item \textit{Energy} = $MM_{e}$(Arrivals)
\end{itemize}
Where $AM$ means arrival model, $MM_{c}$ means mixture model for connection times and $MM_{e}$ means mixture model for the energy charged. `Horizon' is the time horizon for which the data needs to be generated and is an input parameter. Departure time in an EV charging session can be calculated as the sum of its arrival time and connected time.

\subsection{Arrival Models}
\label{sec:arrivalmodel}
    
    Arrivals of EVs in a group of charging stations can be considered as events on a continuous timescale. Supported by our large-scale dataset, we assume that the inter-arrival time of EVs 
    follows an exponential distribution, 
    characterized by a parameter $\lambda$ representing the arrival rate (EVs per time unit).
    This means we can either model the times in between successive events, or rather the number of events in a given time interval:
    \begin{itemize}[topsep=0pt]
        \item \textit{Exponential IAT distribution:} we generate the arrival of the next EV using $\ti = \til + \tdiff$. Where $\ti$ is the time of $i^{th}$ arrival. $\tdiff$ is the time difference between the $i^\textrm{th}$ and ${i-1}^\textrm{th}$ arrivals. The probability density of $\tdiff$ (inter-arrival time, IAT) is exponentially distributed, characterized by $\lambda = f(\textrm{month}, \textrm{daytype}, \textrm{time-of-day})$, where `daytype' is either weekday or weekend.
        \item \textit{Poisson process:} we generate the number of arrivals in a given time slot (\eg slots of 60 minutes). The number of arrivals $N_\textit{arr}$ in a given time slot can be generated using sampling from a Poisson distribution 
        with mean $\lambda \cdot T$ (with $T$ the duration of a timeslot).
    \end{itemize}

        

\begin{figure}[!t]
        \centering
        \includegraphics[width=\columnwidth,height=6cm]{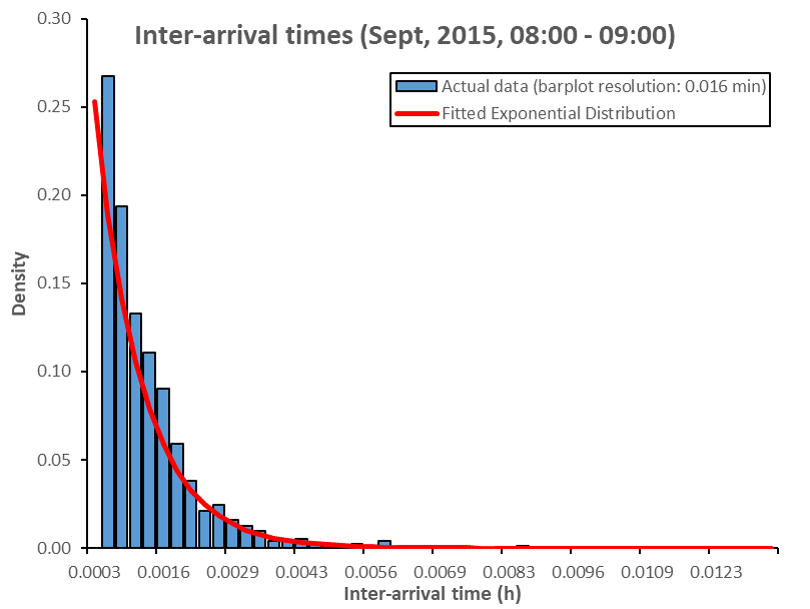}
        \caption{Inter-arrival time probability density for real world data collected by ELaadNL.}
        \label{fig:expdensity}
        \setlength{\belowcaptionskip}{-00pt}
    \end{figure}

    \noindent In \figref{fig:expdensity} we plot density of inter-arrival times where the original data refers to real-world data collected by ELaadNL. A fitted exponential distribution validates this assumption that inter-arrival times are exponentially distributed: Kolmogorov–Smirnov test results for all combinations of month, daytype and time-of-day slots have $p$-value $< 0.05$.
    
    We can model $\lambda$ (in function of month, daytype, time-of-day) using averaging (discontinuous) over each timeslot or rather fitting a continuous function of time (continuous). While fitting continuous curves, capturing the peaks during the day becomes very important. Also, fitted curves could result in negative values of $\lambda$, thus we need to impose lower and upper bounds on it.

    Using the IAT approach, the time of next EV arrival is determined relatively to the time of the previous EV. In case $\lambda$ at a given time-of-day is very low, this implies the next EV arrival may be be generated very late (with a large $\tdiff$), thus skipping several consecutive timeslots. This is problematic if those next timeslots exhibit a much higher $\lambda$, and thus have a high probability of EV arrivals --- which would not be generated because of the very large  $\tdiff$.
    
    The second modeling approach, as a Poisson process, circumvents the above problem, since when certain timeslots have low $\lambda$, we will likely generate 0 arrivals, but still proceed to the immediately next timeslot (with possibly high $\lambda$ to generate possible arrivals there). A potential drawback still is that the variance and mean of the Poisson distribution are equal. In case this assumption does not hold, we adopt a negative binomial distribution instead.

\subsection{Mixture Models for Connected Times and Energy} 
\label{sec:departuremodel}

Aside from arrival events, also EV departure events, as well as the EV charging load (or total energy charged) need to be modeled.
Arrivals are conditional on previous arrivals, obviously, and we model the probability distribution for connected times ($t_\textit{depart} - t_\textit{arr}$) as Gaussian Mixture Models (GMMs). Also for EV charging energy we use GMMs.
As in the arrival generation models, we define and fit different models for each month and time-of-day.




In summary, answering \qref{q:SDGdefination} (Which models can be used in SDG?), we propose
\begin{enumerate*}[(i)]
    \item exponential IAT distribution or Poisson distribution for arrivals per timeslot ($\textit{AM}$),
    \item GMMs for both the duration of EV sessions ($\textit{MM}_c$) and 
    \item their associated energy charged ($\textit{MM}_e$).
\end{enumerate*}

\section{Generating Samples}
\label{sec:training}

After fitting the SDG on a real-world data, we have $AM$, $MM_{c}$ and $MM_{e}$ as parametric models for arrival times, connection times and energy required. For synthetic generation of arrival times, \eg $\lambda$ values can be used to generate $\tdiff$ (and hence a series of arrivals). For connection times and energy required, a sample from the PDFs can be used as a synthetically generated connection time and required energy. This answers \qref{q:generate}, on how we can generate samples.

With our SDG (AM, $MM_{c}$ and $MM_{e}$) can be saved as a separate file. These models will be supplied with the code (that we plan to make publicly available) to generate new synthetic data samples. These models do not include the actual real-world EV session data that the SDG was trained on. Hence, these models can be shared without violating confidentiality constraints. Generated samples from these models can be subsequently used for flexibility analysis, load balancing and other research purposes.

\section{Conclusion and Future Work}
\label{sec:future work}
    Our paper summarized the modeling approach of EV charging sessions. 
    We adopted these models for training a synthetic data generator (SDG) with real-world data, that can thus generate synthetic samples of EV sessions data. 
    We plan to release the source code including the \emph{training} scripts, but also \emph{generation} code (including the settings thereof reflecting the trained model characteristics based on a large-scale real-world dataset) to produce synthetic EV session data reflecting real-world behavior. We believe this fits a strong need of researchers in both academic and industrial settings.
    
    As future work, we aim to propose modeling methods for the time-varying arrival distribution model parameters. 
    Further, we will tackle the following challenges in depth:
    \begin{enumerate*}[label={(\arabic*)}]
    \item Studying the properties of the real-world data with the goal to define evaluations metrics for comparing real-world data with generated samples.
    \item Correctly modeling peaks of arrivals during the day; study effective methods that avoid negative values in continuous $\lambda$ curves. 
\end{enumerate*}

\begin{acks}
This research received funding from the Flemish Government under the “Onderzoeksprogramma Artificiële Intelligentie (AI) Vlaanderen” programme.
\end{acks}

\bibliographystyle{ACM-Reference-Format}
\bibliography{references}



\end{document}